\documentclass[aps,prl,twocolumn,superscriptaddress]{revtex4}
\usepackage{graphicx}
\usepackage{dcolumn}
\usepackage{bm}
\usepackage{graphicx}
\usepackage{subfigure}
\usepackage{amssymb}

\begin{document}

\title{Preferential attachment renders an evolving network of
populations robust against crashes}

\author{Areejit Samal}
\email{samal@mis.mpg.de}
\affiliation{Max Planck Institute for Mathematics in the Sciences,
Inselstr. 22, D-04103 Leipzig, Germany}
\author{Hildegard Meyer-Ortmanns}
\email{h.ortmanns@jacobs-university.de}
\affiliation{School of Engineering and Science, Jacobs University,
P.O. Box 750 561 D-28725 Bremen, Germany}


\begin{abstract}
We study a model for the evolution of chemical species under a combination of population dynamics on a short time scale and a selection mechanism on a longer time scale. Least fit nodes are replaced by new nodes whose links are attached to the nodes of the given network via preferential attachment. In contrast to a random attachment of newly incoming nodes that was used in previous work, this preferential attachment mechanism accelerates the generation of a so-called autocatalytic set after a start from a random geometry and the growth of this structure until it saturates in a stationary phase in which the whole system is an autocatalytic set. Moreover, the system in the stationary phase becomes much more stable against crashes in the population size as compared to random attachment. We explain in detail in terms of graph theoretical notions which structure of the resulting network is responsible for this stability. Essentially it is a very dense core with many loops and less nodes playing the role of a keystone that prevents the system against crashes almost completely.
\end{abstract}
\maketitle


\section{Introduction}

One of the key challenges in understanding the origin of life on prebiotic
earth concerns the creation of non-random and self sustaining entities at
first place from an assumed completely random scenario \cite{Dyson,Kauffman}.
From a naive point of view, it seems extremely unlikely that highly
non-random organizations can evolve from a random start in a  self-organized
manner within a reasonable period of time.
This key question in the origin of life problem can be addressed on
various levels.
Jain and Krishna \cite{JK1} were among the first to illustrate such a
possibility through their mathematical model on the level of graphs
with interacting populations: a highly non-random structure can evolve
from an initial random scenario in a short period of time.
In their evolving network model, the population dynamics on a short time
scale is coupled with a selection mechanism on a longer time scale.
As shown in Ref. \cite{JK1}, it is the appearance of certain non-random
structures in the graph of interacting species together with the
population dynamics assigned to the species that triggers the evolution
of the network towards larger complexity.
These non-random structures turn out to be so-called autocatalytic sets (ACS)
\cite{Dyson,Kauffman,Wach} which are a set of nodes that can
collectively replicate.

In the context of the Jain and Krishna (JK) model \cite{JK1,JK2,JK3,JK4},
an ACS is defined as a set of nodes in the graph where each node
has at least one incoming link from some other node in the same set
\cite{JK1,JK2,JK3}.
One can further divide the nodes of the ACS into two sets:
its core and its periphery (for the precise definition we refer to Ref. \cite{JK3,JK4}).
In the JK model, it is the first occurrence of an ACS after the start from an
initial random graph that triggers the onset of the growth phase.
In the growth phase, there is a dramatic increase in the number of links in the
graph in conjunction with the ACS accreting more and more nodes into it.
The growth phase culminates with the ACS spanning the whole graph. This event marks
the start of the organized phase in which the number of populated nodes
becomes stationary.
The system remains in the organized phase until a crash occurs with a sudden
drop in the number of populated nodes.
After such a crash, the system ends up either in the initial random phase or in
the growth phase from where it starts all over again until it reaches the
organized phase which is followed by an eventual next crash.
In Ref. \cite{JK3,JK4}, different structural
changes in the evolution of the graph were analyzed in detail  that may result in the
crash of the majority of populations in the model.
As it turned out, crashes may happen as a consequence of certain kind of
innovations at an earlier time or of extinction of certain keystone species in
the network. More precisely, crashes are often a result of core-shifts (to be
explained later) \cite{JK3,JK4}.
Recently it was shown that the number of crashes in the model gets
reduced with increase of diversity in the system, diversity in the sense
of larger system size \cite{RVS}.

In this paper, we incorporate a kind of the preferential attachment mechanism \cite{BA,AB}
into the JK model after selection and removal of the weakest node.
We show that preferential attachment
of the  newly added node to existing nodes of higher degree renders the network model
extremely robust against crashes.
So, in our version, we also start with an initial random network and let the populations
of the species evolve according to the same linear population dynamics as the one in the
JK model until it reaches the attractor.
Next, we select the ``worst performing" node or species in the network with the lowest
relative population and eliminate it from the network with all its connections \cite{Bak}.
A new node is then introduced into the network with an average number of incoming and
outgoing links that are attached to existing nodes in the network according to a
preferential attachment rule.
In the context of our model, it means that a new chemical species is more likely to
catalyze (or get catalyzed by) existing nodes which already do catalyze (or get catalyzed by)
several other nodes than those nodes with low connectivity.
The reason why we call the species "chemical" is due to the type of population dynamics.
The model can be argued \cite{JK1} to provide a simplified description of metabolic
networks with substrates with gain and loss terms when they catalyze each other.
The very mechanism of preferential attachment or the ``rich get richer" scheme is certainly
at work in shaping real networks like the World Wide Web (WWW), social and economic
networks \cite{BA,AB}.
In certain cases the preferential attachment mechanism may also explain the origin of
scale-free degree distributions in biological networks \cite{NB}.
In particular, it has been shown that the metabolic networks inside various organisms across
the three domains of life have a scale-free degree distribution with few high-degree
metabolites dominating the overall connectivity of the metabolic networks \cite{JAB,WF}.
Furthermore, according to Ref. \cite{WF} the high-degree metabolites such as ATP and NAD,
which play a central role within the organization of the metabolic network, are also
believed to be the most primitive ones.
Many of these high-degree metabolites also participate in the tricarboxylic acid (TCA)
cycle and glycolysis which are considered to be among the most primitive pathways in
metabolic networks.
This overlap of high-degree metabolites with the list of ancient metabolites has been
argued as an indirect evidence for the role of preferential attachment mechanism in shaping
metabolic networks in the course of evolution.
Similarly, in the Barabasi-Albert model of generating scale-free networks the hubs are
the oldest nodes in the network.
Another strong support for the role of the preferential attachment mechanism in shaping
biochemical networks comes from the recently reported power-law distribution of the
ligand-protein mapping \cite{ligand}.
In Ref. \cite{ligand}, it was shown that there are a few ligands that bind to many
different domains while most ligands bind to few domains.
The highly connected ligands also bind to the most primitive domains in the protein
universe.
This observation again suggests that preferential attachment has certainly played some role
in shaping biochemical networks.

In the JK model, the accelerated growth towards a highly complex and non-random
organization was a result of the occurrence of an ACS by chance along with the selection
of the weakest node.
The reason is that in the growth phase the weakest node cannot belong to the dominant ACS,
the number of nodes and links belonging to the ACS, once it was created, gets amplified.
In our version of the model, in addition to the selection of the weakest node,
the newly added node preferentially attaches to the ACS due to its higher density of
links, and most likely to the nodes forming the core
of the ACS.
This explains why in our version an ACS preferentially
grows along with its core, and core shifts, which are often the main cause for
crashes, are extremely rare.
Furthermore, in our case, the dense architecture of the core goes along with many loops
within the core and therefore multiple sustenance pathways, resulting
in only a few keystone nodes in the graph, so that the structure is much less vulnerable
and becomes extremely stable in the organized phase.
These results should be compared with the error and attack tolerance of scale-free
networks as studied in \cite{reka}.
The error tolerance in scale-free networks is pronounced with respect to random failures
and makes these networks more robust than their random counterparts, but, in contrast to our networks, they are very
vulnerable against intentional attacks due to the presence of their hubs.

The paper is organized as follows.
In section 2, we describe the evolving network model with population dynamics
on a short time scale coupled to selection and  followed by preferential attachment
on a longer time scale.
In section 3, we present the results from simulations of the new model and compare its behavior with that of the JK model.
In secttion 4, we analyze the enhanced stability in terms of structural changes in the graph, generated during the update events.
In section 5 we summarize our results and conclusions.

\section{The Model}

The system consists of $s$ nodes forming a network of directed interactions. The graphical structure
can be completely specified by its adjacency matrix $C  \equiv  (c_{ij}),\ i,j
= 1,\ldots,s$.
A node in the network may be thought to represent a molecular species, the assigned variable gives its concentration.
The element $c_{ij} = 1$ if species $j$  catalyzes the growth of species $i$, and
zero otherwise.
Also, $c_{ii} = 0$ for all nodes $i$, since self replicating species are
excluded from this model.

The equation for the population dynamics of the different molecular species in
the network is then given by
\begin{equation}
\label{equa}
\dot{y}_i\;=\;\sum_{j=1}^{s}C_{ij}\;y_j\;-\;\phi y_i\;.
\end{equation}
Here, $\dot{y}_i$ is the rate of change of the population of species $i$.
The first term on the right hand side of Eq. \ref{equa} can be interpreted as the
positive effect of all species catalyzing species $i$, each one having an
effect proportional to that of its population.
The second term on the right hand side of Eq. \ref{equa} corresponds to a
loss term with $\phi$ as the common death rate for population $y_i$.
In terms of the relative populations $x_i$ with $x_i=y_i/\sum_k y_k$, the Eq.
\ref{equa} can be written as
\begin{equation}
\label{dynamics}
\dot{x}_i\;=\;\sum_{j=1}^{s}c_{ij}\;x_j\;-\;x_i\;\sum_{k,j=1}^{s}c_{kj}\;x_j\;,
\end{equation}
So far the set up of the model is exactly
the same as in the JK model. For a detailed motivation we refer to the original reference \cite{JK1}.

For the dynamics described by Eq. \ref{dynamics}, the system approaches a fixed-point
attractor at which all $x_i$ become time independent constants.
It was shown in \cite{JK3,JK4,Sandeep} that this steady state is just the Perron-Frobenius eigenvector
of the matrix $C$ corresponding to its largest real, so-called Perron-Frobenius
eigenvalue.
For a generic non-negative matrix $C$, there is a unique, global attractor of the
system independent of the choice of initial conditions and stable against perturbations of the
$x_i$.

Next we come to the evolution on a longer time scale.
The first step in the evolving network model involves the generation of a sparse
matrix $C$ that is drawn from the random binomial ensemble with on average $m$
links per node (with $m< 1$).
In the next  step, we use the population dynamics equation given by Eq.
\ref{dynamics} to determine the steady state for the initial random
network. The steady state can be either obtained from the converging dynamics, or,
as mentioned earlier, as the Perron-Frobenius eigenvector of the adjacency matrix $C$
of Eq. \ref{dynamics}.
In order to introduce evolution in the model, selection is imposed in the spirit of Darwin's `Survival of the  fittest'. In this context it is natural what to call least fit:
the species or node with the lowest relative population in
the steady state. This node is removed from the system \cite{Bak} along with all its links from the graph.
If there are more than one least fit nodes, we select one of these candidates at random.

The difference between our model presented here and the JK model lies
in the mechanism by which a new node is added after the removal of the least
fit node from the network.
In the JK model, the newly added node has an average in-degree and out-degree
equal to $m$, it has equal probability of attaching to any of the existing
nodes in the graph.
In our model, the new node shall on average have the same in-degree and
out-degree equal to $m$ as in the JK model, but preferentially
\cite{BA,AB} attaches to an existing node with higher degree.
A new node $i$ will have an an outgoing link and an ingoing link to an existing node $j$ in
the network both with the same probability
\begin{equation}
\label{pref}
\frac{k_j\;m}{\sum_{j=1}^{s} k_j},
\end{equation}
where $k_j$ denotes the total degree of node $j$. The proportionality factor $m$ is used to tune the average (in- and out-) degree by which a new node gets equipped. Here it is chosen such that on average the out- and in degrees are the same as in the random phase and as in the JK model in order to facilitate the comparison of results.
Since the average degree in the simulations
is chosen less than 1 (due to fixing $m$ as $0.25$), in the majority of graph update events the new node will not have any link to the existing nodes, but in some update events it must establish more than one, since otherwise no loops would ever form.
We have checked in our simulations that the average (in- and out-) degrees in the JK model and in our version are approximately equal to $m=0.25$.
Note that the preferential attachment scheme presented here differs from
the standard preferential attachment rule that leads to scale-free degree
distributions in the Barabasi-Albert algorithm \cite{BA,AB} where the
number of incoming links belonging to the new node is fixed in contrast to our scheme where
$m$ gives only an average connectivity of the node over many update events.

The selection and update mechanism together result in a structural
perturbation of the initial network. For the new network
we then determine again the attractor of the population
dynamics Eq. \ref{dynamics} and repeat the selection and update afterwards
to study the interplay between structure
and dynamics during evolution in time. This way two time scales are implemented: The short time scale refers to the population dynamics during convergence to the fixed point, the longer time scale to the time interval between two structural updates, but further time scales are generated dynamically due to the duration of the transient random phase, the transient growth phase, and the interval between two crashes.

In the next section, we present the results from the simulations of our model
in comparison with the JK model.
For simplicity, from now on we will refer to the JK model as the `former
model' and the model presented here with preferential attachment as new ingredient as the
`modified model'.

\section{Results and Discussion}

\begin{figure*}
\centering
\subfigure[]
{
\includegraphics[height=6cm]{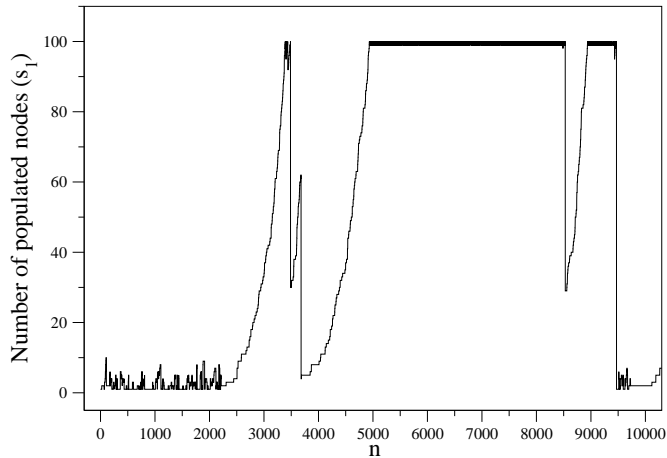}
\label{phasea}
}
\vspace{.3in}
\subfigure[]
{
\includegraphics[height=6cm]{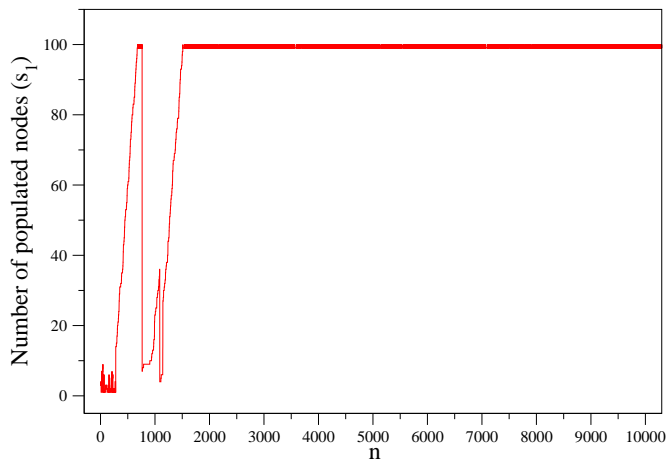}
\label{phaseb}
}
\caption{(Color online) The number of populated nodes, $s_1$, as a function of time
steps, $n$, in (a) the former model and (b) the modified model with number of nodes
$s=100$ and $m=0.25$. To get a better resolution of the initial and the growth phase in the modified model, we plot $s_1$ only for 10000 time steps while
the runs actually extended over $10^5$ steps.
}
\label{phase}
\end{figure*}

\begin{figure*}
\centering
\subfigure[]
{
\includegraphics[height=6cm]{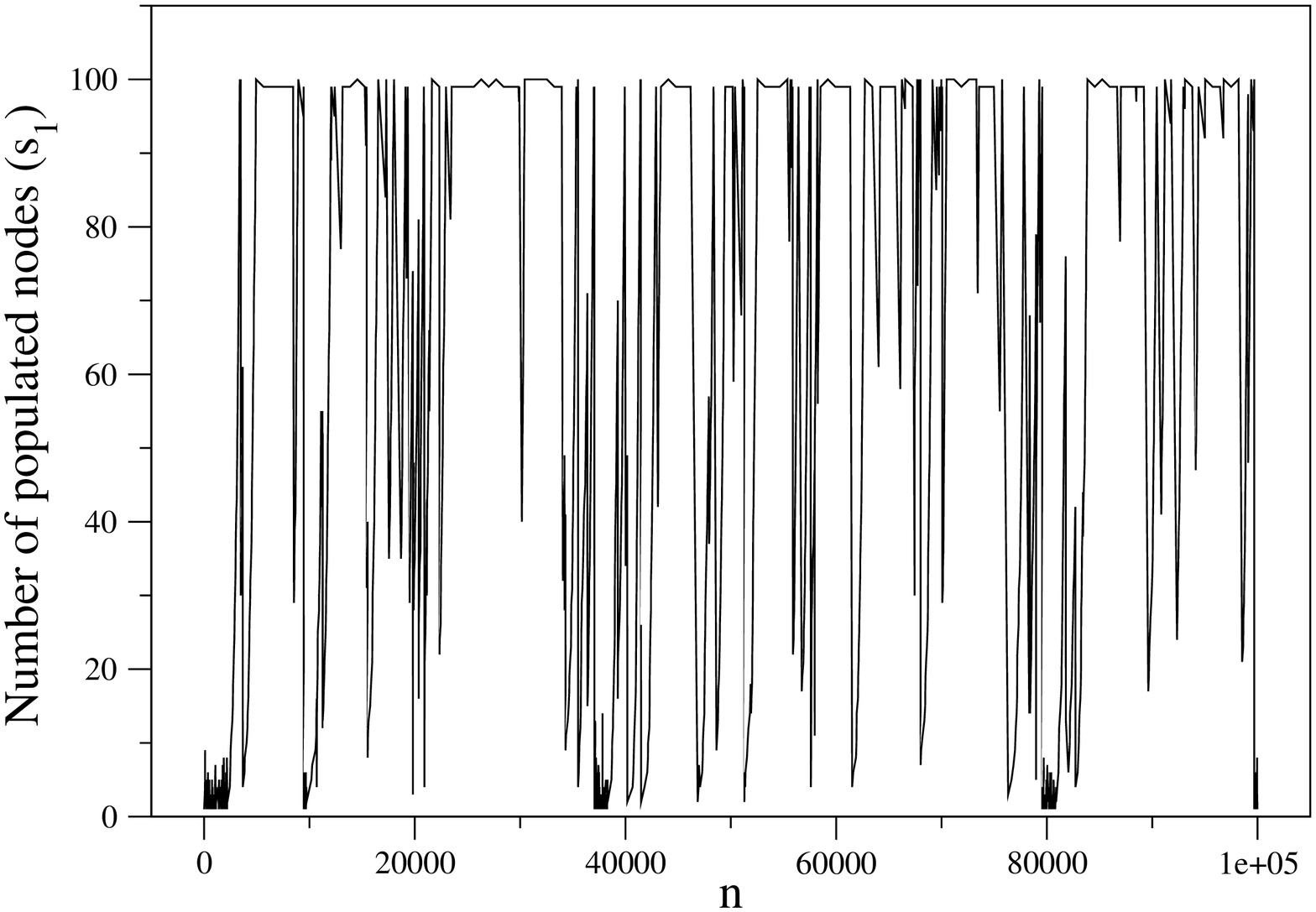}
\label{s1a}
}
\vspace{.3in}
\subfigure[]
{
\includegraphics[height=6cm]{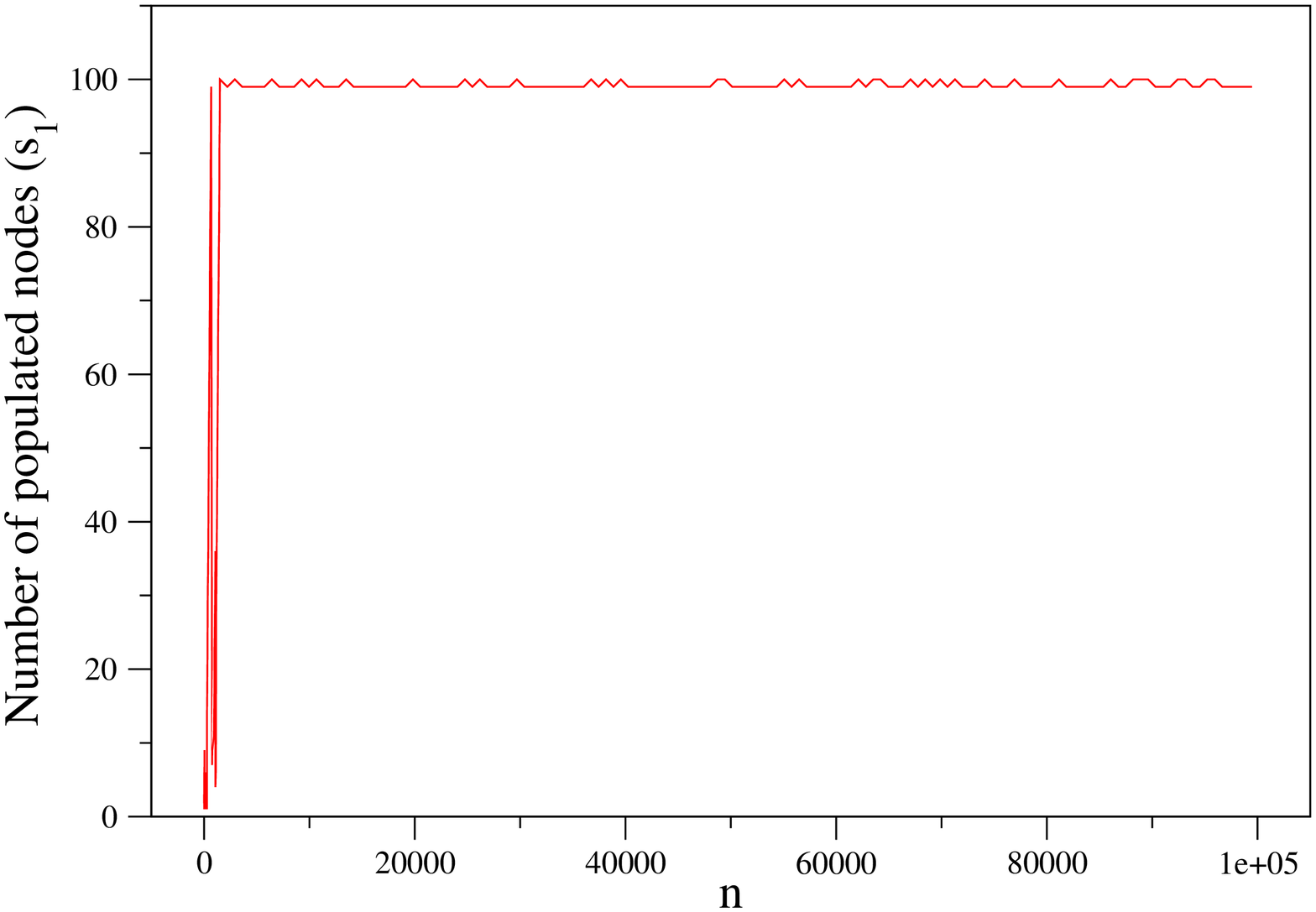}
\label{s1b}
}
\caption{(Color online) The number of populated nodes, $s_1$, as a function of time
steps, $n$, in (a) the former model and (b) the modified model with number of nodes
$s=100$ and $m=0.25$.
The runs shown in part (a) and (b) are the same as those of Fig. \ref{phase} (a) and
(b), respectively, but here we have plotted the value of $s_1$ for each of the $10^5$
time steps.
}
\label{s1}
\end{figure*}

In this section, we present our results from simulations of the former
model and the modified model, and then compare the observed features.
For these simulations, we have chosen the number $s$ of nodes or chemical
species in the network to be equal to 100 and the parameter
$m$ entering the preferential attachment of the new node to be equal to 0.25 in both models
in order to facilitate comparison.
Most of our results presented below are based on data compiled from a
single run of $10^5$ time steps of the former model (shown in Fig.
\ref{phasea} and Fig. \ref{s1a}) and a single run of $10^5$ time steps
of the modified model (shown in Fig. \ref{phaseb} and Fig. \ref{s1b})
with the parameter values $s=100$ and $m=0.25$ in each of them.
In order to check that the runs shown in Fig. \ref{phase} and Fig.
\ref{s1} are actually representative for both models, we have repeated 25
times the run of the former model and the run of the modified
model starting with a different random number seed in each run, but with
the same parameter values $s=100$ and $m=0.25$.
We find that the characteristic behavior observed in the single runs shown in
Fig. \ref{phase} and Fig. \ref{s1} are representative.
For some aspects like the duration of the transient random phase, (i.e.,
the number of time steps starting from the initial random network until the
appearance of the first ACS in the network) and the duration of the first
growth phase, (i.e., the number of time steps from the appearance of the
first ACS until the ACS spans the whole network), we report results averaged
over $1000$ different runs of the former and the modified model,
starting with a different random number seed in each run.
Towards the end of this section, we present results from simulations where we
have varied the number of nodes ($s$) between $s=40, 60, 80$ and $100$ keeping the parameter $m=0.25$ in each case.
Apart from a few test runs we have not varied the value of the parameter $m$, since
it is known for the former model from Ref. \cite{Sandeep} that very small
values of $m$ keep the system in the initial random phase.
Beyond a certain threshold in $m$, the system can transit into
the growth and organized phases in a finite time which in general depends on both values of $m$ and $s$.
The three phases observed during network evolution are explained in detail
below.

\subsection{Three phases are observed during network evolution in both models}

In Fig. \ref{phase}, the number of populated nodes in the graph ($s_1$) is plotted
as a function of time steps ($n$) for the former model and the modified model.
In both models, three phases are observed during network evolution.
The random phase extends from the initial random network until the appearance of
the first ACS in the network.
This is followed by a growth phase where the ACS accretes more and more of the given set of nodes and
grows until an organized structure is built where all $s$ nodes in the network are
populated and are part of the ACS.
The network then continues to stay in the organized phase for some time with at
least $s-1$ nodes populated.
In case of the former model, it is observed that from the organized phase the
network may return to the growth phase or the random phase (an indication for a so-called crash, see below), while in case of the
modified model, the network continues to stay in the organized phase.

\subsection{Preferential attachment mechanism accelerates the creation of the first ACS
and the transition from the growth to the organized phase}

It was found that the first ACS appears after 1107 time steps on an average over $1000$ runs in
the former model while it appears after only $113$ time steps on average in
the modified model, starting from the initial random network.
The duration of the first growth phase was found to be $1600$ time steps on
average in the former model compared to $491$ time steps in the
modified model. All numbers refer to an average over 1000 different runs each
starting with a different random number seed.
Thus, the duration of the initial random phase and the first growth phase are
considerably shorter in the modified model.
This observation can be explained by the graph update scheme.
Starting from a initial random network, the creation of the first ACS is
facilitated by the preferential attachment mechanism in the modified model as
the new node is more likely attaching to existing nodes with higher than with lower degrees.
Further, since the nodes in the ACS have higher degrees compared to nodes
that are not part of the ACS, the new node is more likely attaching to
nodes belonging to the ACS, this obviously accelerates the growth phase in the modified model.

\subsection{Crashes are extremely rare in the modified model}

The most important difference observed between the two models during network
evolution lies in the occurrence of crashes.
A crash in the context of the former model was defined as a single graph
update event in which the number of populated nodes ($s_1$)
drops by more than $s/2$ nodes \cite{JK4}.
During a crash an overwhelming number of nodes
get extinct at the next time step due to some structural change
in the graph.
A comparison in Fig.\ref{s1} shows that crashes are extremely rare in the modified model.

We also compared the number of crashes in both models in a larger data set compiled
from 25 different runs each of $10^5$
time steps for the same parameter values $s=100$ and $m=0.25$
in each run.
A total of 1160 crashes was observed in the 25 runs of $10^5$ time steps each
of the former model, but only a total of 6 crashes were observed in the 25 runs of $10^5$ time
steps each of the modified model.
Further, after the
appearance of the first ACS in the graph, in the modified model the network never returns to the
random phase without any ACS, in contrast to the former case.
Thus, the modified model is extremely robust to crashes which can lead to
the extinction of a majority of species in the network..

Since both models only differ in the graph update scheme, the enhanced
robustness of the modified model must be attributed to the preferential
attachment mechanism.
We now present a detailed analysis of the structure of graphs at different time
steps during network evolution in Fig. \ref{s1}
in order to better understand the origin of robustness to crashes in the modified
model.

\subsection{A typical graph in the organized phase has a much larger
Perron-Frobenius eigenvalue in the modified model}

\begin{figure*}
\centering
\subfigure[]
{
\includegraphics[height=6cm]{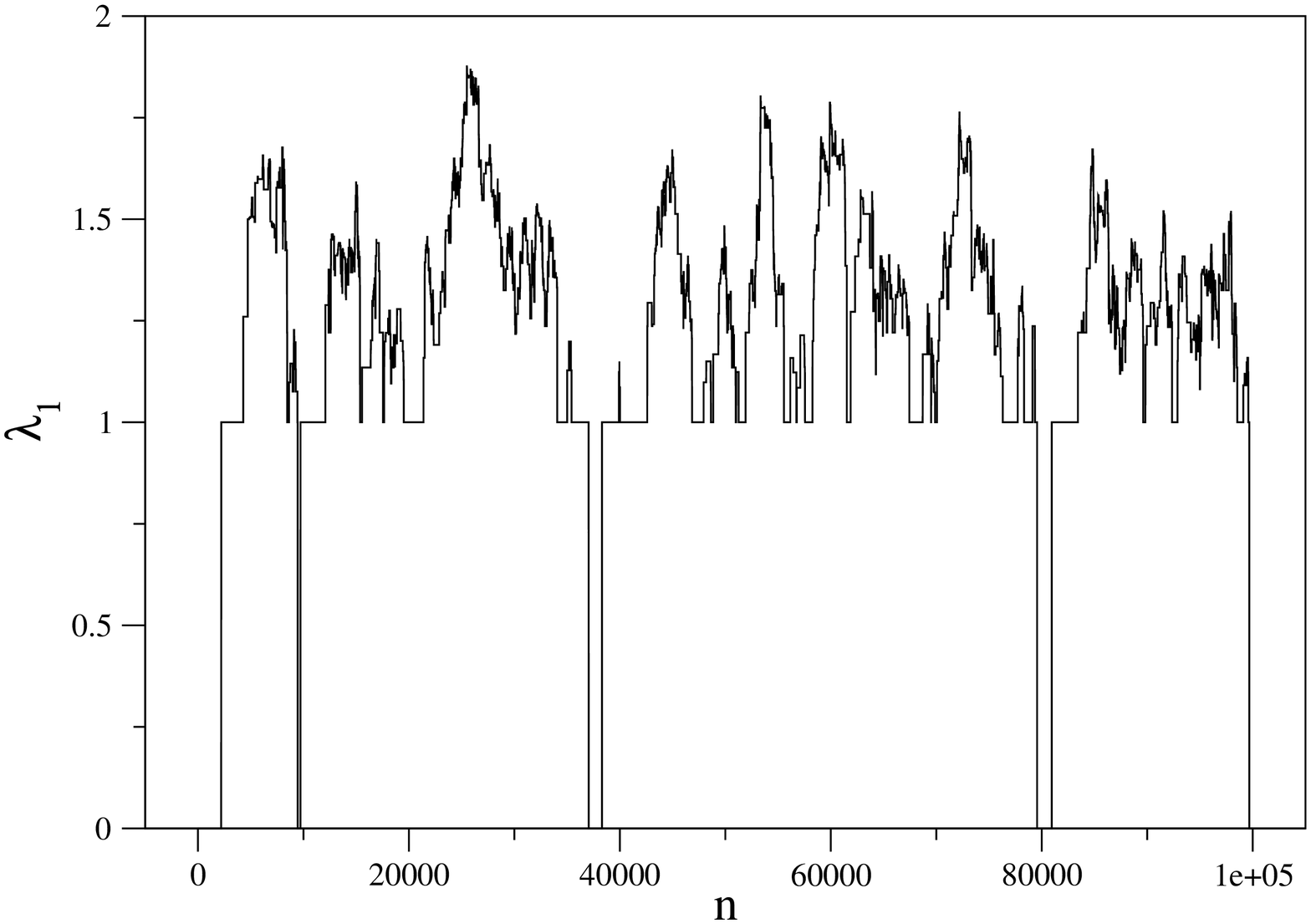}
\label{PFEa}
}
\vspace{.3in}
\subfigure[]
{
\includegraphics[height=6cm]{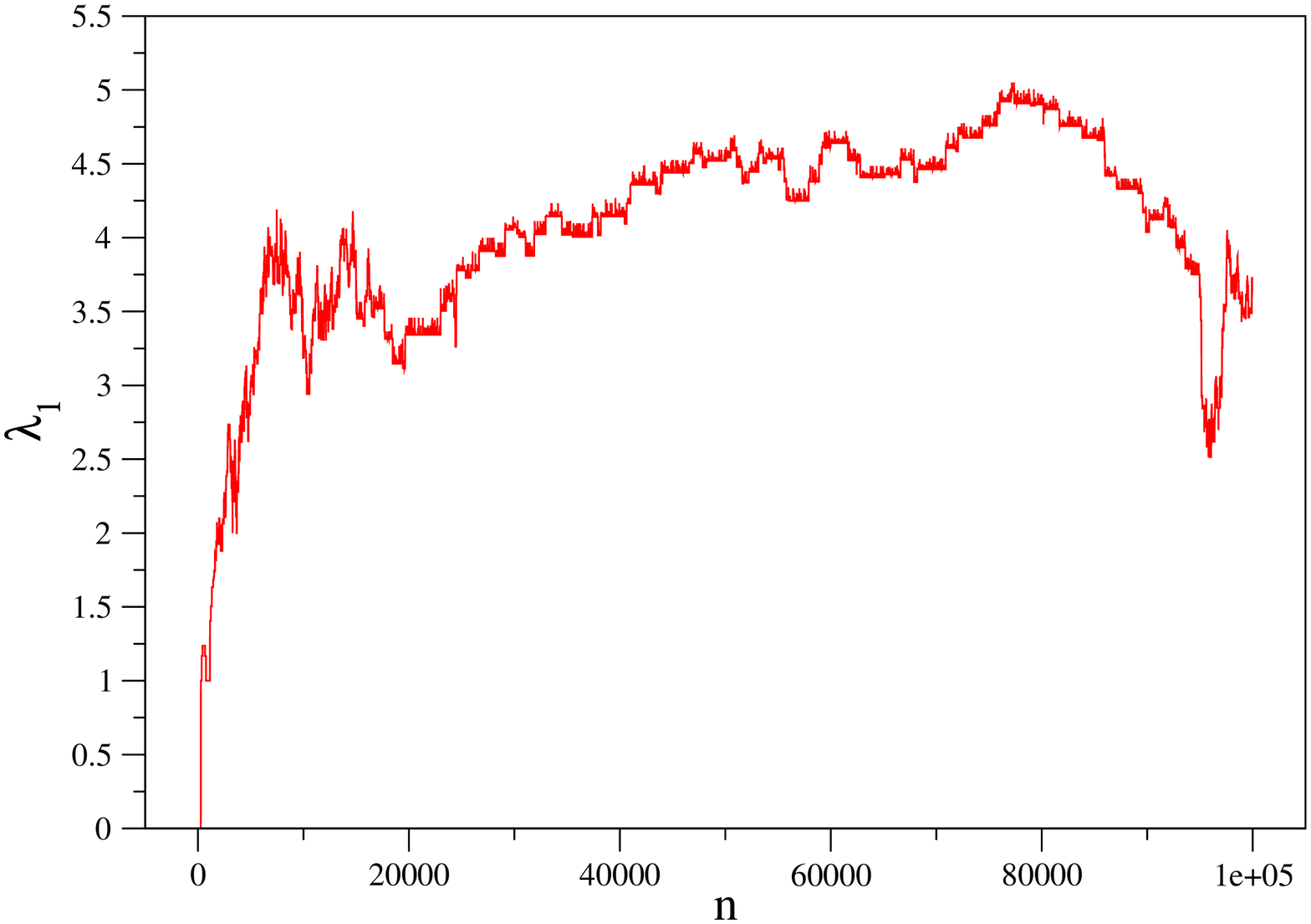}
\label{PFEb}
}
\caption{(Color online) The Perron-Frobenius eigenvalue $\lambda_1$ of the adjacency matrix of a graph, as a function of time $n$, in (a) the former model and (b) the modified
model for the runs shown in Fig. \ref{s1}.
}
\label{PFE}
\end{figure*}

As it was shown in Ref. \cite{JK4}, the Perron-Frobenius
eigenvalue ($\lambda_1$) of the adjacency matrix, assigned to the network and measured as a function of time, is an important measure for elucidating the structural changes in
the graph during the course of network evolution.
Moreover we know from \cite{JK4} that $\lambda_1$ is 0 if the graph has no ACS.
Further, $\lambda_1$ indicates the creation of the first ACS in the graph when it
becomes larger than or equal to $1$.
The appearance of the first ACS leads to the onset of the growth phase which is
characterized by a sudden increase in the number of links in the graph.
In the organized phase, $\lambda_1$ is larger than or equal to $1$ for both models
as there is at least one ACS in the graph in both phases.
We have plotted the time evolution of the Perron-Frobenius eigenvalue ($\lambda_1$) in Fig. \ref{PFE} for both models corresponding to the runs shown in
Fig. \ref{s1}.
By comparing the value of $\lambda_1$ for different graphs in the organized phases (cf. Fig. \ref{PFE}), it is
seen that $\lambda_1$ has a much larger value for a typical graph in the modified
model.
As it was shown in \cite{JK4,Sandeep}, the value of $\lambda_1$ is correlated to the
density of links in the core of the ACS .
Thus the core of the ACS of a typical graph in the modified model is more dense. This renders networks in the organized
phase of the modified model more robust against crashes.

\subsection{The core of a typical graph in the organized phase has many more fundamental
loops in the modified model}

\begin{figure*}
\centering
\includegraphics[height=6cm]{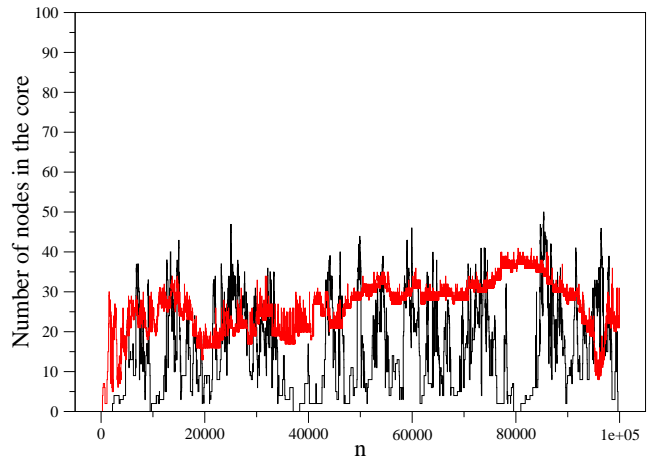}
\caption{(Color online) The number of nodes in the core of the graph as a function of
time steps, $n$, in the former model (black curve) and the modified model (red curve)
for the runs shown in Fig. \ref{s1}.
}
\label{core}
\end{figure*}

\begin{figure*}
\centering
\includegraphics[height=6cm]{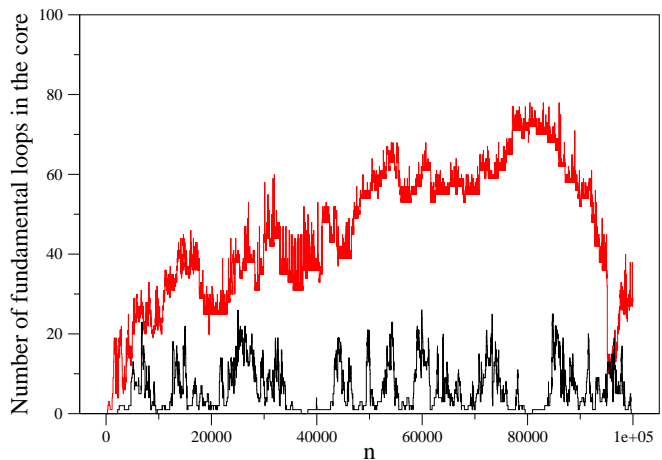}
\caption{(Color online) The number of fundamental loops in the core of the graph as a
function of time steps, $n$, in the former model (black curve) and the modified model
(red curve) for the runs shown in Fig. \ref{s1}.
}
\label{loop}
\end{figure*}

We further explored the structural properties associated with the core of the ACS at
each time step that endows the system with enhanced stability against crashes in the
modified model.
We computed the number of nodes in the core as a measure for its size and the number of fundamental loops
in the core as a function of time for the runs of the two models
shown in Fig. \ref{s1}.
The corresponding data are plotted in Fig. \ref{core} and Fig. \ref{loop}.
Obviously the core size in the former model can become as large as in
the modified model (cf. Fig. \ref{core}), but the number of fundamental loops within the
core is clearly different (cf. Fig. \ref{loop}).
The number of fundamental loops is given by the first Betti
number (or cyclomatic number) which is the number of edges minus the number of nodes
plus the number of connected components, all measured within the core.
The larger number of loops in the core also explains the observed higher
value of $\lambda_1$ in the modified model. It is an indicator of the multiplicity of paths within the core and the reason for its stability.

\begin{figure*}
\centering
\includegraphics[height=6cm]{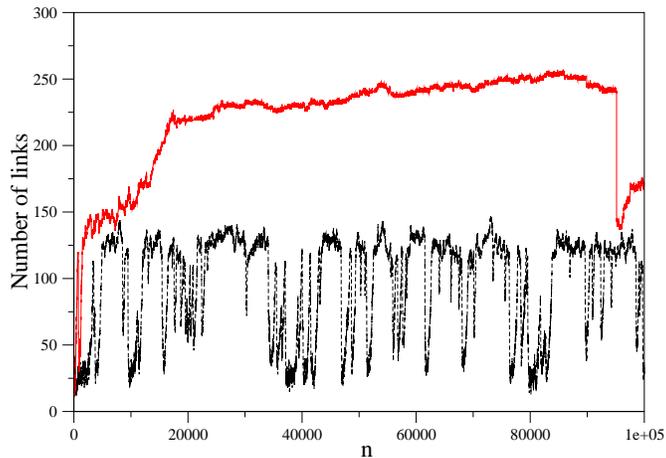}
\caption{(Color online) The number of links in the graph as a function of time steps, $n$,
in the former model (black curve) and in the modified model (red curve) for the runs shown
in Fig. \ref{s1}.}
\label{links}
\end{figure*}

In addition to a denser core, also the total number of links
in the organized phase is larger (cf. Fig. \ref{links}). This is again a consequence of preferential attachment, combined with the "survival" criteria implemented via the population dynamics.

\subsection{Degree distribution and clustering coefficient of a typical graph in the
organized phase}

\begin{figure*}
\centering
\includegraphics[height=6cm]{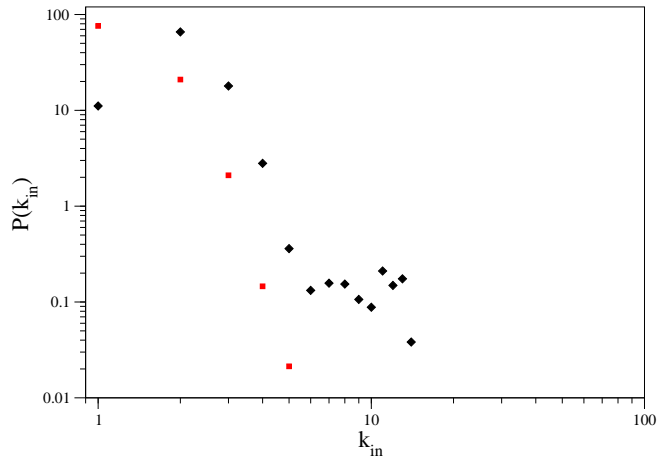}
\caption{(Color online) The in-degree distribution of nodes in the typical graphs of
the organized phase for the former model (black diamonds) and the modified model (red
squares) for the runs shown in Fig. \ref{s1}.}
\label{indeg}
\end{figure*}

\begin{figure*}
\centering
\includegraphics[height=6cm]{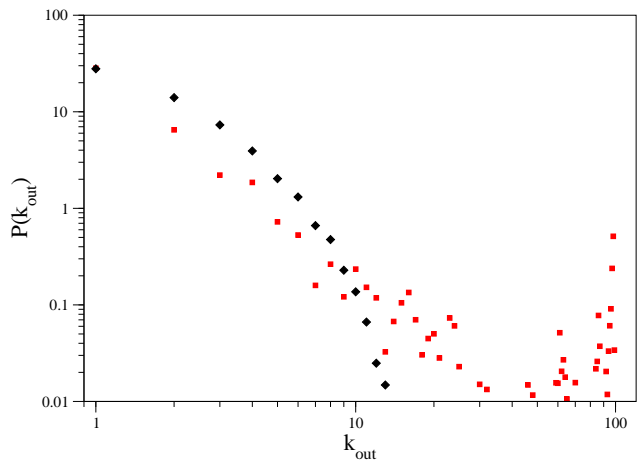}
\caption{(Color online) The out-degree distribution of nodes in the typical graphs of
the organized phase for the former model (black diamonds) and the modified model (red
squares) for the runs shown in Fig. \ref{s1}.}
\label{outdeg}
\end{figure*}

We next study typical graphs in the organized phase of the two models in terms of
standard graph-theoretic measures that characterize the structure of various complex
networks.
The degree distributions for typical graphs in the organized phase of the two
models are shown in Fig. \ref{indeg} and Fig. \ref{outdeg}.
It is clear that both in-degree and out-degree distributions are not exponential
but rather power-like, but it is not possible to uniquely read off the actual values of the power for the degree distributions from our data.
An extension of the degree distribution over more than two orders of magnitude is
also not possible, since the number of nodes ($s$) in our simulations is restricted
to $100$.
The in-degree distribution spans even one order of magnitude
less than the out-degree distribution (cf. Fig. \ref{indeg} and Fig. \ref{outdeg}).
The reason for this asymmetry is a dynamical one, it is not specific to preferential
attachment in the modified model.
The observed asymmetry in the degree distributions can be understood as follows.
The ACS is defined as a set of nodes for which each node has at least one incoming link
from some other node in the set.
The nodes in the ACS can be further divided into two sets: the core and the periphery.
A node in the graph that has an incoming link from a core node but no outgoing link
to any core node will be part of the periphery of the ACS. It will have nonzero
population in the attractor, while, vice versa, a node in the graph that has an outgoing link to a core node but no incoming
link from any core node will not be part of the ACS and will have zero population in
the attractor.
Such a node may get eliminated during selection, so that a formerly connected core node to this node will loose one incoming link in such an event.
Thus, in general, core nodes will have larger out-degrees than in-degrees, as any node that has an
incoming link from a core node is guaranteed to be populated while a node that has
an outgoing link to a core node is not and may get eliminated via selection.

We then computed the clustering coefficient \cite{WS} of the graphs in the organized
phase for the two models.
The average value of the clustering coefficient for the graphs in the organized phase
in the run of the former model was found to be $0.02(9)\pm 0.02$ and $0.7(9)\pm 0.09$ in the modified model.
The much larger value of the clustering coefficient in the
modified model is a mere
reflection of the higher density of links in the core of the graphs.

\begin{figure*}
\centering
\includegraphics[height=8cm]{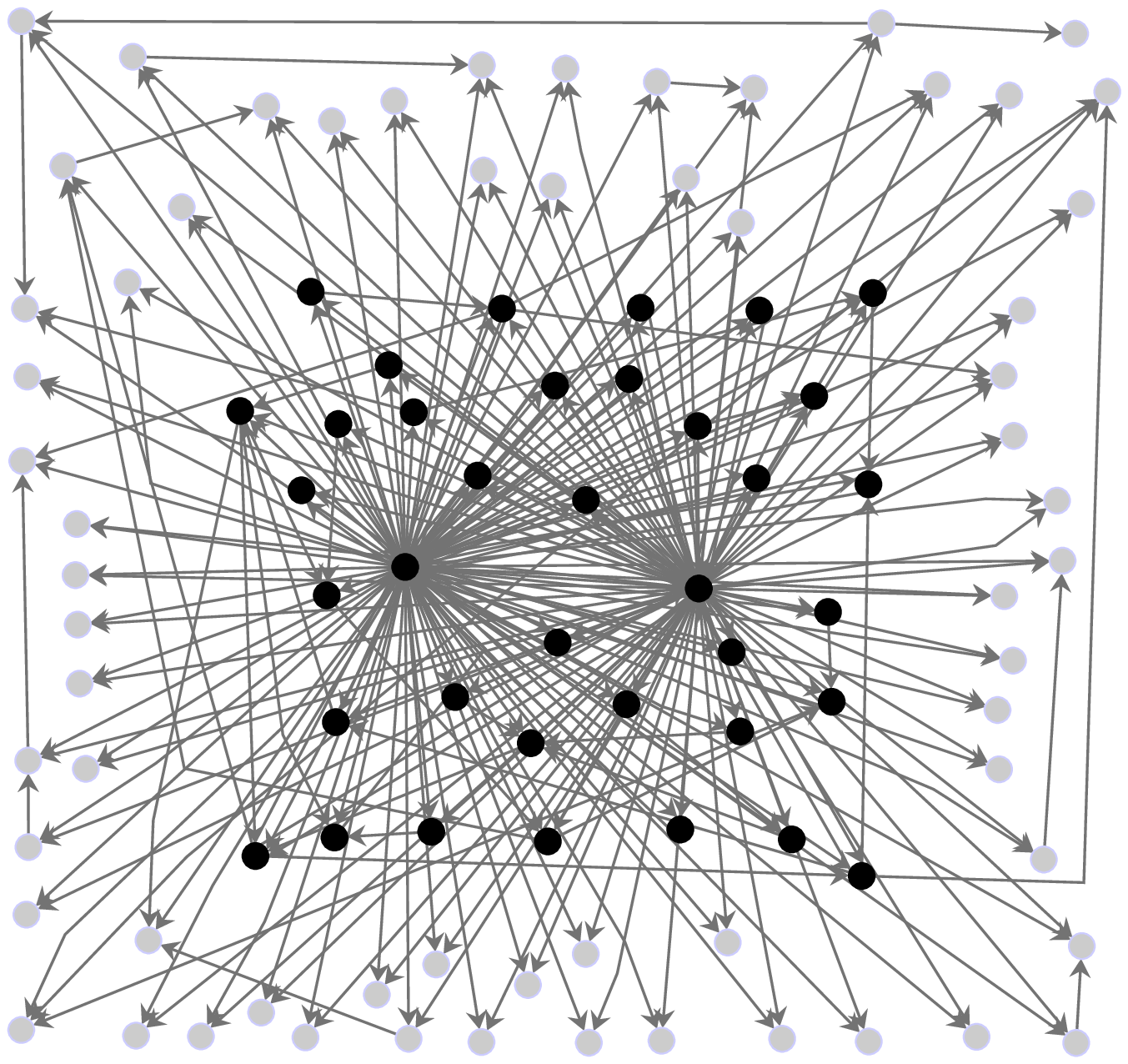}
\caption{The structure of a fully autocatalytic graph in the organized phase of
network evolution in the modified model.
It illustrates the dense core (black nodes) with many loops and the less connected periphery
(grey nodes).
}
\label{g81000}
\end{figure*}

 For illustration we show in Fig. \ref{g81000} the structure of the graph at time step 81000  for
the run of the modified model corresponding to Fig. \ref{s1b} with parameter values
$s=100$ and $m=0.25$.
The graph is fully autocatalytic and all nodes are
populated. The figure nicely shows the dense architecture of the core in the organized phase of modified model with many loops within the core (black nodes).

\subsection{Both diversity and preferential attachment can enhance network robustness}

In Ref. \cite{RVS} it has been observed for the former model that the stability of the model
with respect to crashes increases with an increase in the number of nodes. A larger number of nodes stands for more diversity in the population.
This increase in the stability was attributed to a larger number of cycles in the core of
the graph \cite{RVS}.
We find the same qualitative behavior: an increase of stability of the modified model to crashes with an increase in the number of nodes in the network.
We obtained data on the number of crashes from simulations with $s=40, 60, 80,$ and
$100$ nodes in the graph, keeping the parameter $m=0.25$.
{}From the data in Table 1 we can conclude that the number of crashes increases with a
decrease in the number of nodes.
The data presented here are based on a single run of $10^5$ time steps each, for
$s=40, 60, 80,$ and $100$ nodes with $m=0.25$ for both
models.
These results show that both the preferential attachment mechanism and the increase
in the number of nodes in the network can contribute towards the stability of the
evolving network against crashes.

\begin{table}
\caption{Number of crashes as a function of the number of nodes in the network for the
two models.}
\vspace{1cm}
\centering
\begin{tabular}{|c|c|c|}
\hline
Number of nodes & Number of crashes in & Number of crashes in \\
($s$) & the former model & the modified model \\
\hline
40 & 113 & 23\\
60 & 82 & 6\\
80 & 49 & 2\\
100 & 44 & 1\\
\hline
\end{tabular}
\end{table}

\section{Classification of graph update events into different types
of innovations}

In this section, we present an overview of the types of structural changes
that occur in the graph during update events in the course of network
evolution in the two models.
An innovation in the context of the former model has been defined as a graph
update event in which the relative population of the new added node in the
new attractor is nonzero, i.e, an event in which the new node survives until
the next graph update \cite{JK5,Sandeep}.
Following Ref. \cite{JK5}, we have classified each of the $10^5$ graph update
events in the runs represented in Fig. \ref{s1a} and \ref{s1b}, into the category ``no innovation" and different categories of
innovation.
The results are summarized in Table 2.

\begin{table}
\caption{Classification of the graph update events in the runs of the two
models shown in Fig. \ref{s1} into different categories of innovation.}
\centering
\vspace{1cm}
\begin{tabular}{|l|c|c|}
\hline
Type of Innovation & Former model & Modified model \\
\hline
No innovation & 83670 & 77872\\
Random phase innovation & 112 & 19\\
Incremental innovation & 15130 & 18462\\
Creation of an ACS & 4 & 1\\
Dormant innovation & 92 & 26\\
Core enhancing innovation & 728 & 2970\\
Core reducing innovation & 224 & 649\\
Core shifting innovation & 39 & 1\\
Complete destruction & 1 & 0\\
\hline
\end{tabular}
\end{table}

We now compare the two models based on different categories of innovation.
\begin{itemize}

\item {\bf No innovations:} These are graph update events in which the new node has
a zero relative population in the new attractor.
The data listed in Table 2 show that most graph update events correspond to this category
in the two models as on average the probability of a new node having a link in the network
is much less than unity as the parameter $m$ was set to 0.25 in our simulations.

\item {\bf Random phase innovations:} These are graph update events that occur in the
random phase when there is no ACS in the graph.
The new node has a nonzero relative population in the new attractor
The lower number of random phase innovations in the modified model compared to the
former one can be explained by our findings for the duration of phases: the random phase is
much shorter in the modified model (cf. Table 2).
Also, after the creation of the first ACS in the modified model, the graph contains at
least one ACS in all subsequent time steps.

\item {\bf Incremental innovations:} These are graph update events that occur in the
growth and organized phases when there is an ACS in the graph.
The new node comes in and joins the periphery of the dominant ACS without creating
any new irreducible subgraphs.
The data listed in Table 2 show that the number of such innovations increases in the
modified model.

\item {\bf Creation of an ACS:} These are graph update events that occur in the
random phase when there is no ACS in the graph.
The new node comes in and attaches itself to the existing nodes such that an ACS
is created in the graph.
Further, the new node has nonzero population in the new attractor.
The data of Table 2 show a large number of such events in the former model since the graph there can revert
to the random from the organized phase, so that an ACS can be created more than once in  contrast to the modified model.

\item {\bf Dormant innovations:} These are graph update events that occur in the growth
and organized phases during network evolution.
The new node comes in and attaches to existing nodes such that it creates a new
irreducible subgraph but this irreducible subgraph has a Perron-Frobenius eigenvalue less
than that of the core of the dominant ACS.
The new node has nonzero population in the new attractor. Such irreducible graphs in
the periphery can be the cause for later crashes as a result of competition between different irreducible subgraphs \cite{JK3,JK4}.
Comparing the number of such innovations in the two models, we can see that there are
less such events in the modified model since preferential attachment works against this kind of competition (cf. Table 2).

\item {\bf Core enhancing innovations:} Core enhancing innovations are graph update events that occur in the
growth and organized phases during network evolution.
The new node comes in and attaches to the core of the dominant ACS of the graph at the
previous time step and strengthens it.
The core of the dominant ACS of the updated graph is larger than the core of the graph
at the previous time step.
The set of nodes forming the core of the graph at the present time step has a nonzero
intersection with the set of nodes forming the core of the graph at the previous time
step.
Also, the new node has nonzero population in the new attractor.
The data listed in Table 2 show that in the modified model, there are many more update
events in which the new node strengthens the existing core of the dominant ACS, explaining its enhanced robustness.

\item {\bf Core reducing innovations:} These are graph update events that occur in the
growth and organized phases during network evolution.
The new node comes in and attaches to existing nodes such that the size of the core of
the dominant ACS gets reduced after the update event.
However, the set of nodes forming the core of the graph at the present time step has a
nonzero intersection with the set of nodes forming the core of the graph at the previous
time step.
Also, the new node has nonzero population in the new attractor.
{} From Table 2 we see that there is also a larger number of core reducing events
in the modified model compared to the former one due to preferential attachment to the core with its higher density of links, but these events are compensated by an even
higher number of core enhancing events.

\item {\bf Core shifting innovations:} These are graph update events that can occur in
the growth and organized phases during network evolution.
The deleted node is a keystone node \cite{JK4} and is part of the core of the dominant
ACS.
The new node comes in and attaches to existing nodes such that another dormant irreducible
graph lurking earlier in the periphery of the dominant ACS takes over as the new dominant
ACS of the graph and the new node has nonzero population in the new attractor.
The set of nodes forming the core of the graph at the present time step has no overlap with
the set of nodes forming the core of the graph at the previous time step.
In the modified model, only one such event is observed compared to 39 in the
former model (cf. Table 2).
Thus, core shifting events, which are the main cause for crashes, are extremely rare when
the preferential attachment mechanism is introduced.

\item {\bf Complete destruction:} These are graph update events that occur in the growth
and organized phases during network evolution when there is an ACS in the graph.
The deleted node is a keystone node in the graph and the new node comes in and attaches
to the existing nodes such that no ACS is left in the new graph.
However, the new node has a nonzero relative population in the new attractor.
Only one such event is observed in the original model and none in the modified model among
the $10^5$ graph update events in both models (cf. Table 2).
Such events are extremely rare in both cases.
\end{itemize}

\section{Summary and Conclusions}

In this paper, we have incorporated the preferential attachment mechanism into
the graph update scheme of an earlier model for evolving networks
\cite{JK1,JK2,JK3}.
In the modified model presented here, three phases are observed during network
evolution similar to those observed in the former model.
However, the initial random phase and the first growth phase are much shorter
now as compared to the former model.
Furthermore, crashes are extremely rare in the modified model.
We have studied in detail how the incorporation of a preferential attachment
mechanism into the graph update scheme renders the structure of the network
robust against crashes during the course of evolution.
We find that the core of the ACS becomes highly dense in terms of the number
of links during the course of network evolution in the modified model.
The ACS of the graph including its core are endowed with much more stability due to a large number of fundamental loops in the
core as a result of the preferential attachment. Along with that, there are
multiple pathways in the core, only very few
keystone nodes and seldom core shift events.
Coming back to the key question in the origin of life problem on how to generate
complex non-random structures in a reasonable time when starting from a completely
random scenario: here we have presented an efficient mechanism in the attachment of new nodes that not only accelerates the creation
of complex structures, but also renders them stable against crashes when all other parameters are kept fixed.


\subsection*{Acknowledgments}
We would like to thank Sanjay Jain for discussions and suggestions
and Sandeep Krishna for making the computer program available to
simulate the former model.
One of us (AS) would also like to thank Ravi Mehrotra and Vikram
Soni for discussions, and Rainer Kleinrensing for technical help.
He gratefully acknowledges a postdoctoral fellowship from the Max
Planck Institute for Mathematics in the Sciences (MPI-MIS), Leipzig,
and support from the International Center for Transdisciplinary Studies
(ICTS) at Jacobs University, where part of the project was done.



\end{document}